\documentclass[10pt]{IEEEtran}
\usepackage{amsmath,amsfonts}
\usepackage{color}
\usepackage{array}
\usepackage{algorithmic}
\usepackage[linesnumbered,ruled,vlined]{algorithm2e}
\usepackage[caption=false,font=normalsize,labelfont=sf,textfont=sf]{subfig}
\usepackage{textcomp}
\usepackage{stfloats}
\usepackage{url}
\usepackage{times}
\usepackage{verbatim}
\usepackage{graphicx}
\usepackage[nocompress]{cite}
\usepackage{longtable}
\usepackage[utf8]{inputenc}
\usepackage[T1]{fontenc}
\usepackage{lipsum}
\usepackage{multirow}
\usepackage{mathtools}
\usepackage{titlesec}
\usepackage{setspace}
\usepackage[utf8]{inputenc}
\usepackage[linesnumbered,ruled,vlined]{algorithm2e}
\usepackage{amsmath, amssymb}

\makeatletter
\renewcommand\paragraph{\@startsection{paragraph}{4}{\z@}%
                                    {1.3ex plus 1.5ex minus 0.2ex}%
                                    {0.7ex plus .2ex}%
                                    {\normalfont\normalsize\itshape}}
\renewcommand\subparagraph{\@startsection{subparagraph}{5}{\z@}%
                                       {1.3ex plus 1.5ex minus 0.2ex}%
                                       {0.7ex plus .2ex}%
                                       {\normalfont\normalsize\itshape}}
\makeatother

\newcommand{\argmax}{\operatornamewithlimits{argmax}}

\begin{document}
\pagenumbering{gobble}

\linespread{0.97}
\title{Study of Robust Multiuser Scheduling and Power Allocation in Cell-Free MIMO Networks}

\author{Saeed Mashdour$^{\star}$, André R. Flores $^{\star}$, Shirin Salehi $^{\star\star}$, Rodrigo C. de Lamare $^{\star,\dagger}$, Anke Schmeink $^{\star\star}$ 

\thanks{The authors are with the Centre for Telecommunications Studies, Pontifical Catholic University of Rio de Janeiro, Brazil 
$^{\dagger}$, University of York, United Kingdom, $^{\star\star}$ Chair of Information Theory and Data Analytics, RWTH Aachen University, 52062 Aachen, Germany $^{\ddagger}$
smashdour@gmail.com, andre.flores@esp.puc-rio.br, delamare@puc-rio.br, \{shirin.salehi, anke.schmeink\}@inda.rwth-aachen.de}
\thanks{This work was supported by CNPQ, CPQD and FAPERJ.}}

\maketitle

\setstretch{1.08}

\begin{abstract}
This paper introduces a robust resource allocation framework for the downlink of cell-free massive multi-input multi-output (CF-mMIMO) networks to address the effects caused by imperfect channel state information (CSI). In particular, the proposed robust resource allocation framework includes a robust user scheduling algorithm to optimize the network's sum-rate and a robust power allocation technique aimed at minimizing the mean square error (MSE) for a network with a linear precoder. Unlike non-robust resource allocation techniques, the proposed robust strategies effectively counteract the effects of imperfect CSI, enhancing network efficiency and reliability. Simulation results show a significant improvement in network performance obtained by the proposed approaches, highlighting the impact of robust resource allocation in wireless networks.

\end{abstract}


\section{Introduction}
Cell-free massive multi-input multi-output (CF-mMIMO) networks represent a major advancement in wireless communications that builds on the success of massive multi-input multi-output (MIMO) systems \cite{marzetta,mmimo,wence}. This architecture is defined by the deployment of numerous access points (APs) distributed through the whole network, to serve multiple user equipments (UEs) across the network areas \cite{ngo2015cell}. Such a design significantly enhances the overall system performance by mitigating fading and interference effects of cellular networks \cite{interdonato2019ubiquitous, bjornson2020scalable,llridd, ssettumba2024centralized,oclidd}. 

Although CF-mMIMO networks significantly improve signal quality and system capacity, in order to maximize the potential of these systems, efficient resource allocation is essential for ensuring equitable service and balancing the load among the numerous APs and UEs \cite{nayebi2016performance, buzzi2017cell,tds,jpba,mashdour2024clustering}. An appropriate resource allocation might involve sophisticated strategies for precoding \cite{siprec,gbd,wlbd,mbthp,robprec,bbprec,rcfprec,lrcc}, power control \cite{lrpa,rapa,rscf}, and user scheduling, all aimed at enhancing network performance \cite{demir2021foundations, d2021improving}.  

Effective user scheduling in cell-free networks is essential to ensure that interference is minimized and network throughput is maximized especially in scenarios where the number of UEs exceeds the number of APs \cite{mashdour2022MMSE-Based, dimic2005downlink}. Power allocation also plays a crucial role in resource management by appropriate distribution of the power among APs to enhance the network capacity and energy efficiency \cite{van2016joint}. However, in practical scenarios, the effectiveness of resource allocation strategies in CF-mMIMO networks is significantly impacted by the imperfect knowledge of channel state information (CSI) due to time variability, feedback delays and errors, and overhead. Therefore, the design of robust resource allocation algorithms against imperfect CSI is fundamental to ensure reliable network performance \cite{ bjornson2015optimal}. 

Addressing the challenges of imperfect CSI has recently attracted a great deal of research. In \cite{bashar2018robust}, the optimization of user scheduling in the uplink of massive MIMO systems has been addressed in order to reduce the channel estimation overhead by employing the COST 2100 channel model. This method minimizes the need for comprehensive channel information while maintaining the efficiency of user scheduling. The study in \cite{chang2017energy} investigates an energy-efficient resource allocation algorithm for a WPT-enabled multi-user massive MIMO system, focusing on optimizing antenna selection, power allocation, and time allocation to enhance energy efficiency, and shows that accounting for imperfect CSI increases the algorithm's robustness. \cite{rong2005robust} explores robust beamforming-type linear receiver techniques aimed at improving joint space-time decoding and interference rejection. It adopts a worst-case performance optimization strategy, emphasizing that the adaptations of minimum variance methods are particularly tailored to address the difficulties posed by imperfect CSI at the receiver. The development of precoders for a multi-user MIMO (MU-MIMO) system has been investigated in \cite{vaezy2018energy} in two scenarios: one with imperfect CSI and known channel error statistics, and another with norm-bounded error lacking statistical characterization. It presents an energy-efficient precoder design for MU-MIMO systems, aimed at improving robustness in the presence of imperfect CSI.

In this work, we introduce a robust resource allocation framework against imperfect CSI for the downlink of CF-mMIMO networks where the linear Minimum Mean Square Error (MMSE) precoder is used. We propose a robust multi-user scheduling algorithm based on a worst-case optimization \cite{rspa}. Subsequently, we present a robust power allocation technique which employs a robust gradient descent power allocation (RGDPA) algorithm, integrating channel estimation errors into the optimization process for efficient power distribution. Numerical results illustrate the excellent performance of the proposed techniques.

The organization of this paper is as follows: In Section \ref{SYS.mod}, system model and the sum-rate expression is presented for CF network. In Section \ref{Robust.PA}, the robust resource allocation framework is presented including a robust multiuser scheduling and a robust power allocation method which incorporates the channel estimation matrix. In Section \ref{Simul}, the results are presented and Section \ref{conclud} concludes the paper.

The notations used are as follows: $\mathbf{I}_{n}$ for an $n\times n$ identity matrix, $\mathcal{CN}$ for the complex normal distribution, and superscripts $^{T}$, $^{\ast}$, and $^{H}$ for transpose, complex conjugate, and Hermitian (conjugate transpose), respectively. Union and set difference are denoted by $\mathcal{A}\cup \mathcal{B}$ and $\mathcal{A}\setminus \mathcal{B}$. $\mathbb{C}^{M\times N}$ represents complex matrices of size $M\times N$, while $Tr(.)$ indicates matrix trace. $\textup{diag}(\mathbf{x})$ turns vector $\mathbf{x}$ into a diagonal matrix; $\textup{diag}(\mathbf{X})$ extracts diagonal elements from $\mathbf{X}$. The expectation is represented by $\mathbb{E}[.]$, and $\mathbf{X}_{mn}$ specifies the element at the $m$th row and $n$th column of $\mathbf{X}$.

\section{System model and non-robust resource allocation} \label{SYS.mod}

We examine the downlink of a CF-mMIMO network setup comprising $M$ single-antenna APs collaborating to serve $K$ single antenna UEs. We assume that the number of UEs significantly exceeds the total number of AP antennas ($K\gg M$). This underscores the need for multiuser scheduling to guarantee optimal system performance. Consequently, during each resource block transmission, the network arranges a selection of $n \leq M$ UEs. Moreover, the network has the capability to distribute power among the $n$ scheduled users to improve the sum-rates.

\subsection{CF Network and Sum-Rate}

We represent the channel matrix between APs and UEs by $\mathbf{G}=\hat{\mathbf{G}}+\tilde{\mathbf{G}}\in \mathbb{C}^{M\times n}$, where $\hat{\mathbf{G}}$ is the channel estimation matrix and $\tilde{\mathbf{G}}$ is the channel estimation error. The channel coefficient between the $m$th AP antenna and UE $k$ shown by $g_{mk}=\left [ \mathbf{G} \right ]_{m,k}$ is modeled as follows:
\begin{equation} \label{eq.gI}
\begin{split}
    g_{{mk}} = & \hat{g}_{mk}+\tilde{g}_{mk}\\
    =&\sqrt{1-\alpha}\sqrt{\beta_{mk}}h_{mk} + \sqrt{\alpha}\sqrt{\beta_{mk}}\tilde{h}_{mk},
\end{split}
\end{equation}
where $\hat{g}_{mk}$ represents the estimated channel coefficient and $\tilde{g}_{mk}$ denotes the estimation error, \( 0<\alpha<1  \) serves as a parameter addressing CSI imperfections, balancing the influence of the small-scale channel estimation coefficient \( h_{mk} \) and its error \( \tilde{h}_{mk} \), and $\beta_{mk}$ is the large-scale fading (LSF) coefficient. \( h_{mk} \) is characterized by independent and identically distributed (i.i.d.) random variables (RVs), remaining constant within a coherence interval but varying independently across different intervals \cite{ngo2017cell}, and following a complex Gaussian distribution with zero mean and unit variance. Similarly, \( \tilde{h}_{mk} \) has a complex Gaussian distribution with zero mean and unit variance, and is independent of \( h_{mk} \). We consider \( \mathbf{P} \in \mathbb{C}^{M\times n}\) as a linear MMSE precoder. Thus, the downlink signal is modeled as follows:
\begin{equation} \label{CF.sig}
\begin{split}
\mathbf{y}=&\sqrt{\rho_{f}}\mathbf{G}^T\mathbf{P}\mathbf{x}+\mathbf{w}\\
=&\sqrt{\rho_{f}}\hat{\mathbf{G}}^T\mathbf{P}\mathbf{x}+\sqrt{\rho_{f}}\tilde{\mathbf{G}}^T\mathbf{P}\mathbf{x}+\mathbf{w},
\end{split}
\end{equation}
where $\mathbf{x} = [x_{1}, \ldots, x_{n}]^{T}$ is the zero mean symbol vector and follows a complex normal distribution, $\mathbf{x} \sim \mathcal{CN}(\mathbf{0}, \mathbf{I}_{n})$, and the additive noise vector $\mathbf{w} = [w{1}, \ldots, w_{n}]^{T}$ is distributed as $\mathbf{w} \sim \mathcal{CN}(0, \sigma_{w}^{2}\mathbf{I}_{n})$. We assume Gaussian signaling, where the elements of $\mathbf{x}$ are statistically independent and also independent from noise and channel coefficients. The received signal includes an error component due to imperfect CSI, as described in the received signal model. Given these conditions, along with Gaussian signaling and the statistical independence of $\mathbf{x}$ and $\mathbf{w}$, the sum-rate expression is given by
\begin{equation}\label{eq:RCF}
    SR=\log_{2}\left (  \det\left [\mathbf{R}_{CF}+\mathbf{I}_K  \right ]\right ),
\end{equation}
where the covariance matrix $\mathbf{R}_{CF}$ is described by
\vspace{-2mm}
\begin{equation}\label{eq:RCF_1}
    \mathbf{R}_{CF}=\rho _{f} \hat{\mathbf{G}}^{T}\mathbf{P}\mathbf{P}^{H}\hat{\mathbf{G}}^{\ast }\left ( \rho _{f}\tilde{\mathbf{G}}^{T}\mathbf{P}\mathbf{P}^{H}\tilde{\mathbf{G}}^{\ast } +\sigma _{w}^{2}\mathbf{I}_K\right )^{-1}.
\end{equation}

\subsection{Non-Robust Resource Allocation}

In CF networks, resource allocation is crucial for improving the performance of the system. As the number of UEs $K$ is larger than the total number of AP antennas $M$, it is necessary to consider the user scheduling to serve an efficient subset of UEs at any given time. In this regard, we consider the following optimization problem: 
\begin{equation} \label{rop1}
    \begin{aligned}
& \underset{\mathcal{S}_{n}}{\text{max}}~SR\left ( \mathcal{S}_{n} \right ) \\
& \text{subject to} \ \left \| \mathbf{P}\left ( \mathcal{S}_{n} \right ) \right \|_{F}^{2}\leq P,
\end{aligned}
\end{equation}
where ${S}_{n}$ is the selected set of UEs, and $P$ is the upper limit of the signal covariance matrix ${Tr}\left [ \mathbf{C}_{\mathbf{x}}\right ]\leq P$ and $\mathbf{P}\left ( \mathcal{S}_{n} \right )\in \mathbb{C}^{M\times {n}}$ is the related  precoding matrix.

For power allocation which aims to enhance the sum-rate performance, we consider the signal in (\ref{CF.sig}), comprising a desired component alongside terms related to imperfect CSI and noise. Incorporating the power allocation matrix, we rewrite the received signal as follows:
\begin{equation} \label{yc2}
\begin{split}
\mathbf{y}&=\sqrt{\rho _{f}}\hat{\mathbf{G}}^T\mathbf{W}\mathbf{D}\mathbf{x}+\sqrt{\rho _{f}}\tilde{\mathbf{G}}^T\mathbf{P}\mathbf{x}
+\mathbf{w}.
\end{split}
\end{equation}
We have considered the precoding matrix in the desired term as $\mathbf{P}=\mathbf{W}\mathbf{D}$, which is a multiplication of the normalized MMSE precoder matrix $\mathbf{W}\in\mathbb{C}^{NL \times {n}}$  and the power allocation matrix $\mathbf{D}\in\mathbb{C}^{{n} \times {n}}$ that is described by
\begin{multline}
\mathbf{D}=\begin{bmatrix}
\sqrt{p_{1}} & 0 & \cdots  & 0\\ 
 0& \sqrt{p_{2}} & \cdots &0 \\ 
 \vdots & \vdots  &\cdots   & \vdots \\ 
 0&0  &\cdots   & \sqrt{p_{{n}} }
\end{bmatrix}=\textup{diag}\left ( \mathbf{d} \right )\\
, \mathbf{d}
=\left [ \sqrt{p_{1}} \ \sqrt{p_{2}} \ \cdots \  \sqrt{p_{{n}}} \right ]^T
.\end{multline}
Since minimizing the mean squared error (MSE) is equivalent to the sum-rate maximization as described in \cite{mashdour2022MMSE-Based}, we consider the objective function as the MSE between the transmitted signal and the received signal. Thus, the optimization problem is formulated as
\begin{equation} \label{optmse.1}
\begin{aligned}
& \underset{\mathbf{d}}{\text{min}}~\mathbb{E}\left [ \varepsilon  \right ] \\
& \text{subject to}\ \left \| \mathbf{W} \textup{diag}\left ( \mathbf{d} \right ) \right \|^{2}\leq P,
\end{aligned}
\end{equation}
where the error is
\begin{equation}\label{err.def}
    \varepsilon =\left \| \mathbf{x}-\mathbf{y} \right \|^{2}.
\end{equation}
 Imperfect CSI complicates resource allocation in wireless networks. In optimization, the impact of imperfect CSI is clear, as it can result in inferior choices and a decline in performance. Thus, the given optimization problems in (\ref{rop1}) and (\ref{optmse.1}), will require to be updated to consider the effect of imperfect CSI. Consequently, we design a robust version of the given problems that can counteract the uncertainties introduced by CSI errors. 

\section{Robust User Scheduling and Power Allocation} \label{Robust.PA}

In the proposed robust resource allocation technique, we first consider the robust multiuser scheduling when equal power loading (EPL) is considered. Subsequently, we proceed with a robust power allocation algorithm based on a gradient descent power allocation technique.

\subsection{Robust user scheduling}

To robustify the multiuser scheduling in (\ref{rop1}) against imperfect CSI, we devise a worst-case scenario and propose the robust multiuser scheduling optimization problem as
\begin{equation} \label{R.O.P}
\begin{aligned}
& \underset{\mathcal{S}_{n}}{\text{max}}~\underset{\tilde{\mathbf{G}}_a }{\text{min}}~SR\left ( \mathcal{S}_{n} \right ) \\
& \text{subject to} \\
& \quad \left \| \mathbf{P}_a\left ( \mathcal{S}_{n} \right ) \right \|_{F}^{2}\leq P,\\
& \quad \ 
\beta\leq \left\| \tilde{\mathbf{g}}_{k}\right\|^{2}\leq \beta_0 , \forall k \in \mathcal{S}_{n}.
\end{aligned}
\end{equation}
The problem in (\ref{R.O.P}) seeks to maximize the sum-rate under worst-case scenarios, incorporating constraints on the channel estimation error, represented by $\tilde{\mathbf{G}}_a$ for the aggregate and $\tilde{\mathbf{g}}_{k}$ for individual users, bounded by $\beta$ and $\beta_0$. Unlike \eqref{rop1}, this approach uses a dual-layer optimization to balance sum-rate optimization with robustness against the worst channel conditions.

Since the problem in (\ref{R.O.P}) is a complicated non-convex problem, we use a sub-optimal strategy inspired by the clustered
enhanced subset greedy (C-ESG) multiuser scheduling in \cite{mashdour2022enhanced} and propose the robust clustered enhanced greedy (RC-ESG) algorithm. With RC-ESG, we first implement a robust greedy user scheduling approach to select a set of UEs which maximizes the sum-rate. In the robust greedy method, we use a robust optimization problem as follows to select the first selected UE:
\begin{equation} \label{rop2}
\begin{aligned}
& \underset{k\in \mathcal{K}}{\textup{argmax}} ~\left ( \underset{\tilde{\mathbf{g}}_{k}}{\min} \left ( \mathbf{g}_{{k}}^{H}\mathbf{g}_{{k}} \right )\right )\\ \
& \text{subject to}\\
&\ \beta\leq \left\| \tilde{\mathbf{g}}_{k}\right\|^{2}\leq \beta_0 , \forall k\in \mathcal{K}.
\end{aligned}
\end{equation}
Treating the chosen UE as the first member of the UE set, we iteratively update this set by adding a UE that, when combined with the UE set of the previous iteration, yields the highest sum-rate. This process continues until $n$ UEs are selected, forming the UE set obtained through a robust greedy approach. After selecting the first UE set, we identify a UE as the worst UE $k_{wr}$ from this first selected set that could adversely affect the overall system performance under the worst estimation error using the following optimization problem:
\begin{equation} \label{rop3}
\begin{aligned}
& \underset{k\in \mathcal{S}_{n}(j)}{\textup{argmin}} ~\left ( \underset{\tilde{\mathbf{g}}_{k}}{\max} \left ( \mathbf{g}_{{k}}^{H}\mathbf{g}_{{k}} \right )\right )\\ \
& \text{subject to}\\
&\ \beta\leq \left\| \tilde{\mathbf{g}}_{k}\right\|^{2}\leq \beta_0, \forall k\in \mathcal{S}_{n}(j).
\end{aligned}
\end{equation} 
Thereafter, from the remaining UEs $\mathcal{K}_{\text{re}}$ apart from the selected set by the robust greedy approach, we select a UE as best UE $k_{b}$ in a manner that is robust against the CSI imperfection based on an optimization problem similar to the problem in (\ref{rop2}). By replacing the worst UE from the first set by the best UE, we obtain the second UE set. We iterate this process until we obtain a number of UE sets such that we can approach the UE set obtained by the exhaustive search method while a substantial computational complexity is saved. Among the obtained sets, the UE set with the highest sum-rate is the desired set.

In order to solve the optimization problems such as problems given in (\ref{rop2}) and (\ref{rop3}), we optimize the CSI imperfection level $\alpha$ using gradient descent or gradient ascent techniques depending on the convexity of the objective $\mathbf{g}_{{k}}^{H}\mathbf{g}_{{k}}$ with respect to $\alpha$. The proposed RC-ESG technique is summarized in Algorithm \ref{alg:robust_alg}.

\SetKwInput{KwInput}{Input}
\SetKwInput{KwOutput}{Output}
\SetKwInput{KwInitialization}{Initialization}
\SetKw{KwGoTo}{go to}
\SetKwComment{Comment}{/* }{ */}

\begin{algorithm}
\caption{RC-ESG Algorithm}\label{alg:robust_alg}
\Begin{
     $j=1$, $l = 1$\;
    Set $\beta, \beta_0$ using system feedback\;
    $k_{1}=\argmax_{k\in \mathcal{K}} \left (  \min_{\tilde{\mathbf{g}}_{k}} \left ( \mathbf{g}_{{k}}^{H}\mathbf{g}_{{k}} \right )  \right )$ , \quad 
    $\text{s.t. } \beta\leq \left\| \tilde{\mathbf{g}}_{k}\right\|^{2}\leq \beta_0 , \forall k \in \mathcal{K}$
    and denote $k_{1}=\mathcal{U}_{1}$ and $SR_{\text{MMSE}}(\mathcal{U}_{1})$\;
}

\While{$l < n$}{
    $l = l+1$\;
    Set current $\beta, \beta_0$\;
    
    $k_{l}=\argmax_{k\in(\mathcal{K} \setminus \mathcal{U}_{l-1})} \left (SR\left ( \mathcal{U}_{l-1} \cup {k} \right ) \right )$    \;
    Set $\mathcal{U}_{l}=\mathcal{U}_{l-1} \cup {k_{l}}$, denote $SR(\mathcal{U}_{l})$\;
    \If{$SR(\mathcal{U}_{l}) \leq SR(\mathcal{U}_{l-1})$}{
        \KwGoTo end\;
    }

}
$\mathcal{S}_{n}(j)=\mathcal{U}_{l}$\;
$\mathcal{K}_{\text{re}}(j)=\mathcal{K}\setminus \mathcal{S}_{n}(j)$\;
  $k_{wr}(j)=\underset{k\in \mathcal{S}_{n}(j)}{\textup{argmin}}\left ( \underset{\tilde{\mathbf{g}}_{k}}{\max} \left ( \mathbf{g}_{{k}}^{H}\mathbf{g}_{{k}} \right )\right )$,  \quad 
    $\text{s.t. } \beta\leq \left\| \tilde{\mathbf{g}}_{k}\right\|^{2}\leq \beta_0 , \forall k\in \mathcal{S}_{n}(j)$;
    
 $k_{b}(j)=\underset{k\in \mathcal{K}_{\text{re}}(j)}{\textup{argmax}}~\left ( \underset{\tilde{\mathbf{g}}_{k}}{\min} \left ( \mathbf{g}_{{k}}^{H}\mathbf{g}_{{k}} \right )\right )$,  \quad 
    $\text{s.t. }\beta\leq \left\| \tilde{\mathbf{g}}_{k}\right\|^{2}\leq \beta_0 , \forall k\in \mathcal{K}_{\text{re}}(j)$;
    
\For{$j=2$ \KwTo $K-n+1$}{

$\mathcal{S}_{n}(j)=\left (\mathcal{S}_{n}(j-1)\setminus k_{wr}(j-1) \right )\cup k_{su}(j-1)$\;
        $\mathcal{K}_{\text{re}}(j)=\mathcal{K}_{\text{re}}(j-1)\setminus k_{su}(j-1)$\;
    $k_{wr}(j)=\underset{k\in \mathcal{S}_{n}(j)}{\textup{argmin}}\left ( \underset{\tilde{\mathbf{g}}_{k}}{\max} \left ( \mathbf{g}_{{k}}^{H}\mathbf{g}_{{k}} \right )\right )$,  \quad 
    $\text{s.t. } \beta\leq \left\| \tilde{\mathbf{g}}_{k}\right\|^{2}\leq \beta_0 , \forall k\in \mathcal{S}_{n}(j)$;
    
 $k_{b}(j)=\underset{k\in \mathcal{K}_{\text{re}}(j)}{\textup{argmax}}~\left ( \underset{\tilde{\mathbf{g}}_{k}}{\min} \left ( \mathbf{g}_{{k}}^{H}\mathbf{g}_{{k}} \right )\right )$,  \quad 
    $\text{s.t. }\beta\leq \left\| \tilde{\mathbf{g}}_{k}\right\|^{2}\leq \beta_0 , \forall k\in \mathcal{K}_{\text{re}}(j)$;
   
 Compute: $SR(\mathcal{S}_{n}(j))$\;
   
}

$\mathcal{S}_{n}^{d}=\argmax_{\mathcal{S}_{n} \in \mathcal{S}_{n}(m)} \left \{ SR\left ( \mathcal{S}_{n} \right ) \right \}$\;
Precoding of $\mathcal{S}_{n}^{d}$\;
\end{algorithm}

\subsection{Robust Power Allocation} \label{RPA1}

To robustify the given optimization problem in (\ref{optmse.1}) against imperfect CSI, we incorporate the channel estimation error into the optimization problem by defining a robust optimization problem as follows:
 \begin{equation} \label{R.PA.c}
\begin{aligned}
& \underset{\mathbf{d}}{\text{min}}~\mathbb{E}\left [ \varepsilon \mid \hat{\mathbf{G}} \right ] \\
& \text{subject to} \\
& \quad \left\| \mathbf{W}\text{diag}\left( \mathbf{d} \right) \right\|^{2}\leq P.
\end{aligned}
\end{equation}
This method considers the uncertainty in channel estimation on the basis of the expected error $\varepsilon$ on the estimated channel matrix $\hat{\mathbf{G}}$. Consequently, the optimization problem recognizes the existence of estimation inaccuracies, resulting in a power allocation strategy that is robust to imperfect CSI.
 
 We use the definition of the error in (\ref{err.def}) and employ the property ${Tr}\left ( \mathbf{A}+\mathbf{B} \right )={Tr}\left ( \mathbf{A} \right )+{Tr}\left ( \mathbf{B} \right )$, where $\mathbf{A}$ and $\mathbf{B}$ are matrices with the same dimensions, and rewrite the error equation as follows:
 \begin{equation} \label{eq:trerr}
 \begin{split}
          \varepsilon & ={Tr}\left (\mathbf{x}^{H}\mathbf{x}  \right )-\left( {Tr}(\mathbf{x}^{H}\mathbf{w}  \right )-{Tr}\left (\mathbf{w}^{H}\mathbf{x}  \right )+{Tr}\left (\mathbf{w}^{H}\mathbf{w}  \right )+\\ 
       &{Tr}\left (\rho _{f}\mathbf{x}^{H}\textup{diag}\left ( \mathbf{d}\right )\mathbf{W}^{H}\hat{\mathbf{G}}^*\hat{\mathbf{G}}^T\mathbf{W}\textup{diag}\left ( \mathbf{d}\right )\mathbf{x}  \right )+\\
       &{Tr}\left (\rho_{f}\mathbf{x}^{H}\mathbf{P}^{H}\tilde{\mathbf{G}}^*\tilde{\mathbf{G}}^T\mathbf{P}\mathbf{x}\right )-\\
       &{Tr}\left (\sqrt{\rho _{f}}\mathbf{x}^{H}\hat{\mathbf{G}}^T\mathbf{W}\textup{diag}\left ( \mathbf{d}\right )\mathbf{x}  \right )-\\
       & {Tr}\left (\sqrt{\rho _{f}}\mathbf{x}^{H}\textup{diag}\left ( \mathbf{d}\right )\mathbf{W}^{H}\hat{\mathbf{G}}^*\mathbf{x}  \right )-\\
       &{Tr}\left ( \sqrt{\rho _{f}}\mathbf{x}^H {\tilde{\mathbf{G}}}^T\mathbf{P}\mathbf{x} \right )-{Tr}\left ( \sqrt{\rho _{f}}\mathbf{x}^{H}\mathbf{P}^{H}\tilde{\mathbf{G}}^*\mathbf{x} \right )+\\
       &{Tr}\left ( \rho_{f}\mathbf{x}^{H}\textup{diag}\left ( \mathbf{d}\right)\mathbf{W}^{H}\hat{\mathbf{G}}^*{\tilde{\mathbf{G}}}^T\mathbf{P}\mathbf{x}\right )+\\
   &{Tr}\left ( \rho_{f}\mathbf{x}^{H}\mathbf{P}^{H}\tilde{\mathbf{G}}^*\hat{\mathbf{G}}^T\mathbf{W}\textup{diag}\left ( \mathbf{d}\right)\mathbf{x} \right )+\\
       & {Tr}\left (\sqrt{\rho _{f}}\mathbf{x}^{H}\textup{diag}\left ( \mathbf{d}\right )\mathbf{W}^{H}\hat{\mathbf{G}}^*\mathbf{w}  \right )+\\
       &{Tr}\left (\sqrt{\rho _{f}}\mathbf{w}^{H}\hat{\mathbf{G}}^T\mathbf{W}\textup{diag}\left ( \mathbf{d}\right )\mathbf{x}  \right )+\\
       &{Tr}\left ( \sqrt{\rho _{f}}\mathbf{w}^{H}\tilde{\mathbf{G}}^T\mathbf{P}\mathbf{x} \right )+{Tr}\left ( \sqrt{\rho _{f}}\mathbf{x}^{H}\mathbf{P}^{H}\tilde{\mathbf{G}}^*\mathbf{w} \right ).
\end{split}
\end{equation}
Considering $\mathbf{x} \sim \mathcal{CN}(\mathbf{0}, \mathbf{I}_{n})$, $\mathbf{w} \sim \mathcal{CN}(0, \sigma_{w}^{2} \mathbf{I}_{n})$, and elements of the $\tilde{\mathbf{G}}$ as zero-mean random variables that are mutually independent, we can derive the expectation of the conditioned error as follows:
\begin{equation} \label{exp.err.con}
  \begin{split}
  \mathbb{E}\left [ \varepsilon \mid \hat{\mathbf{G}} \right ]=
    &n+n\sigma _{w}^{2}+\\
    &{Tr}\left (\rho _{f}\textup{diag}\left ( \mathbf{d}\right )\mathbf{W}^{H}\hat{\mathbf{G}}^*\hat{\mathbf{G}}^T\mathbf{W}\textup{diag}\left ( \mathbf{d}\right ) \right )+\\&{Tr}\left (\rho_{f}\mathbf{P}^{H}\tilde{\mathbf{G}}^*\tilde{\mathbf{G}}^T\mathbf{P}\right )-\\
    &{Tr}\left (\sqrt{\rho _{f}}\hat{\mathbf{G}}^T\mathbf{W}\textup{diag}\left ( \mathbf{d}\right ) \right )-\\
    &{Tr}\left (\sqrt{\rho _{f}}\textup{diag}\left ( \mathbf{d}\right )\mathbf{W}^{H}\hat{\mathbf{G}}^* \right ).
    \end{split}
  \end{equation}
  

 To obtain the derivative of the conditional error expectation with respect to the power allocation factors, we use $\frac{\partial \text{Tr}(\mathbf{AB})}{\partial \mathbf{A}} = \mathbf{B} \odot \mathbf{I}$, where $\mathbf{A}$ denotes a diagonal matrix, $\odot$ is the Hadamard product, and the cyclic property of the trace operator. Thus, we obtain: 
\begin{equation} \label{first.or}
 \begin{split}
     \frac{\partial \mathbb{E}\left [ \varepsilon \mid \hat{\mathbf{G}} \right ]  }{\partial \mathbf{d}}=&2\rho _{f}\left ( \mathbf{W}^{H}\hat{\mathbf{G}}^*\hat{\mathbf{G}}^T\mathbf{W}\textup{diag}\left ( \mathbf{d}\right ) \right )\odot \mathbf{I}-\\
     &2\sqrt{\rho _{f}}Re\left \{ \mathbf{W}^{H}\hat{\mathbf{G}}^*\odot \mathbf{I} \right \}.
     \end{split}
 \end{equation}
Since we consider the MSE as a convex function, the conditional error expectation is convex with respect to the power allocation factors. Thus, in order to solve the optimization problem in (\ref{R.PA.c}), we use a gradient descent technique such that in each iteration, we  update the power allocation coefficients as follows:
 \begin{equation}
     \mathbf{d}\left ( i \right )=\mathbf{d}\left ( i-1 \right )-\lambda \frac{\partial \mathbb{E}\left [ \varepsilon \mid \hat{\mathbf{G}} \right ] }{\partial \mathbf{d}}\bigg|_{\mathbf{d} = \mathbf{d} \left ( i -1\right )},
\end{equation}
where $i$ is the iteration index, and $\lambda$ is the positive step size and $\frac{\partial \mathbb{E}\left [ \varepsilon \mid \hat{\mathbf{G}}_a \right ] }{\partial \mathbf{d}}$ is obtained using (\ref{first.or}). After the iterations, to satisfy the transmit power constraint $\left \| \mathbf{W} \textup{diag}\left ( \mathbf{d} \right ) \right \|^{2}=\left \| \mathbf{P} \right \|^{2}\leq P$, the power allocation coefficients are scaled as follows: 
 \begin{equation}
 \mathbf{d}=\eta\mathbf{d}, \ \eta =\sqrt{\frac{{Tr}\left ( \mathbf{P}\mathbf{P}^{H} \right )}{{Tr}\left ( \mathbf{W}\textup{diag}\left ( \mathbf{d}.\mathbf{d} \right )\mathbf{W}^{H} \right )}}.
\end{equation}

 Accordingly, we can outline the proposed robust gradient descent power allocation (RGDPA) technique in Algorithm \ref{alg:RPA}.

\begin{algorithm}
\caption{RGDPA Algorithm}\label{alg:RPA}

\KwInput{$\mathbf{G}$, $\mathbf{P}$, $\mathbf{W}$, $\lambda$, $\mathbf{d}\left ( 1 \right )$, $\textup{I}_{D}$ }

    \For{$i = 2$ \KwTo $\textup{I}_{D}$}{
        $\frac{\partial \mathbb{E}\left [ \varepsilon \mid \hat{\mathbf{G}} \right ]  }{\partial \mathbf{d}} = 2\rho _{f}\left ( \mathbf{W}^{H} \hat{\mathbf{G}}^*\hat{\mathbf{G}}^T\mathbf{W}\textup{diag}\left ( \mathbf{d}\right ) \right )\odot \mathbf{I}-
      2\sqrt{\rho _{f}}Re\left \{ \mathbf{W}^{H}\hat{\mathbf{G}}^*\odot \mathbf{I} \right \}$ \;
        $\mathbf{d}\left ( i \right ) = \mathbf{d}\left ( i-1 \right )-\lambda \frac{\partial \mathbb{E}\left [ \varepsilon \mid \hat{\mathbf{G}}_a \right ]  }{\partial \mathbf{d}}\bigg|_{\mathbf{d} = \mathbf{d} \left ( i -1\right )}$\;} 
  ${\mathbf{d}_d} =\mathbf{d} ( {\bf I}_D )$; 
    \If{${Tr}\left ( \mathbf{W}\textup{diag}\left ( \mathbf{d}_d.\mathbf{d}_d \right )\mathbf{W}^{H} \right ) \neq {Tr}\left ( \mathbf{P} \mathbf{P}^{H} \right )$}{
            $\eta =\sqrt{\frac{{Tr}\left ( \mathbf{P}\mathbf{P}^{H} \right )}{{Tr}\left ( \mathbf{W}\textup{diag}\left ( \mathbf{d}_d.\mathbf{d}_d \right )\mathbf{W}^{H} \right )}}$\;
            $\mathbf{d}_d=\eta \mathbf{d}_d$\;
        }
\end{algorithm}

When in Algorithm \ref{alg:RPA} we consider non-conditioned error expectation $\mathbb{E}\left [ \varepsilon \right ]$, since $\hat{\mathbf{G}}_a$ is also treated as zero-mean random variable and independent from other variables, the error expectation and its first derivative with respect to the power allocation factors are obtained as follows, respectively:
\begin{equation} \label{exp.err}
  \begin{split}
  \mathbb{E}\left [ \varepsilon \right ]=
    &n+n\sigma _{w}^{2}+ {Tr}\left (\rho_{f}\mathbf{P}^{H}\tilde{\mathbf{G}}^*\tilde{\mathbf{G}}^T\mathbf{P}\right )+\\
    &{Tr}\left (\rho _{f}\textup{diag}\left ( \mathbf{d}\right )\mathbf{W}^{H}\hat{\mathbf{G}}^*\hat{\mathbf{G}}^T\mathbf{W}\textup{diag}\left ( \mathbf{d}\right ) \right ),
    \end{split}
  \end{equation}

\begin{equation} \label{der1.err.exp} 
 \begin{split}
     \frac{\partial \mathbb{E}\left [ \varepsilon \right ]  }{\partial \mathbf{d}}=&2\rho _{f}\left ( \mathbf{W}^{H}\hat{\mathbf{G}}^*\hat{\mathbf{G}}^T\mathbf{W}\textup{diag}\left ( \mathbf{d}\right ) \right )\odot \mathbf{I}.
     \end{split}
 \end{equation}
Thus, the non-robust gradient descent power allocation (GDPA) algorithm is regarded as similar to Algorithm \ref{alg:RPA}, but the power factors are updated in each iteration using $\frac{\partial \mathbb{E}\left [ \varepsilon \right ]  }{\partial \mathbf{d}}$ as obtained in (\ref{der1.err.exp}).

\section{Simulation Results} \label{Simul} 

We consider a CF-mMIMO network including $M=64$ APs and $K=32$ UEs. Using C-ESG user scheduling we select $n=16$ UEs. We examine systems with both perfect and imperfect CSI cases. For robust resource allocation scenarios, we establish bounds on channel estimation error power (denoted by $\beta$, $\beta_0$, $\beta_1$, and $\beta_2$) to yield an imperfect channel factor range of $0.05 \leq \alpha \leq 0.3$. This range is intentionally chosen to maintain the channel within a controlled boundary, avoiding significant deviation toward unreliability or an overly idealized perfect channel state.

\begin{figure}
	\centering
		\includegraphics[width=.85\linewidth]{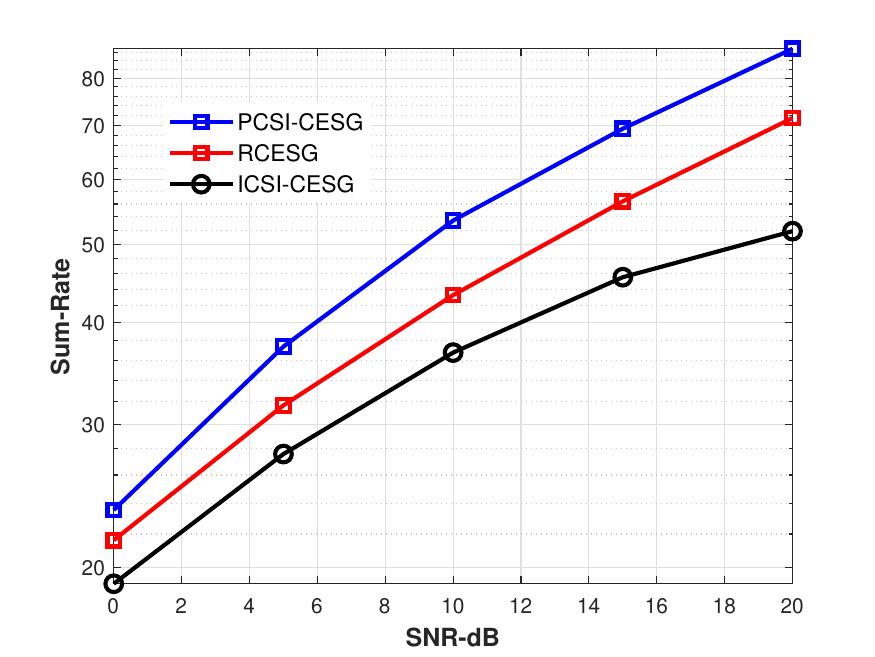}
	\caption{\small{Comparison of the C-ESG and RC-ESG in PCSI and ICSI CF networks when EPL is used and  $\alpha=0.15$ for ICSI case, $M=64$, $K=32$ and $n=16$.}}
	\label{fig:fig1}
\end{figure}

Following this, we implement the robust multiuser scheduling strategy developed. In comparison of the C-ESG effectiveness with perfect CSI (PCSI) and imperfect CSI (ICSI), as well as the performance of RC-ESG with EPL, we fix $\alpha = 0.15$ for the network with ICSI. The outcomes, as illustrated in Fig.~\ref{fig:fig1}, clearly show that RC-ESG outperforms C-ESG for the networks with ICSI, approaching the performance observed with PCSI. However, it should be noted that RC-ESG is somewhat more complex than C-ESG, due to the iterations required to solve the optimization problems in (\ref{rop2}) and (\ref{rop3}).

\begin{figure}
	\centering
		\includegraphics[width=.85\linewidth]{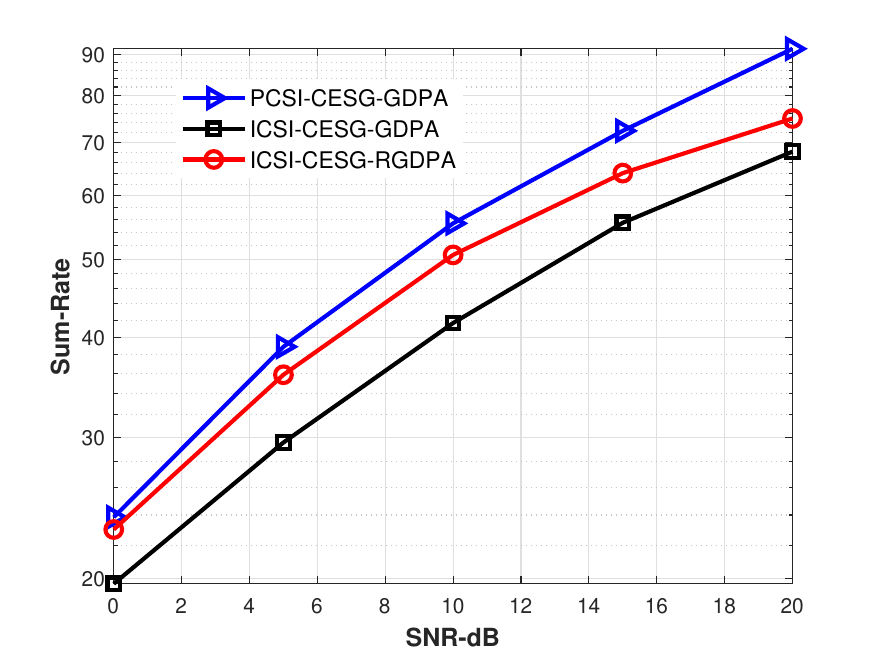}
	\caption{\small{Comparison of the RGDPA and GDPA power allocation techniques when C-ESG user scheduling is used, $\alpha=0.15$ for ICSI case, $M=64$, $K=32$ and $n=16$.}}
	\label{fig:fig2}
\end{figure}

In Fig.~\ref{fig:fig2}, we present a comparison between the proposed GDPA and RGDPA power allocation algorithms, assuming the CESG multiuser scheduling algorithm is employed within the network. This comparison which is performed by setting the uncertainty level for ICSI as  $\alpha=0.15$, reveals that the RGDPA algorithm achieves superior performance to GDPA, as it approaches the results obtained in PCSI networks.  It is important to highlight that RGDPA requires a slightly higher level of complexity than GDPA, primarily due to the extra calculations required for (\ref{first.or}) as compared to (\ref{der1.err.exp}) in the GDPA.

\section{Conclusions} \label{conclud}

In this work, we developed a robust resource allocation framework for the downlink of the CF network with imperfect CSI. The proposed strategy employs a worst-case robust user scheduling algorithm to maximize sum-rates and a robust power allocation strategy based on minimizing conditional MSE. Simulations confirm that our methods significantly enhance the network robustness against the effects of imperfect CSI and leads to substantial performance improvements.

 
\bibliographystyle{IEEEbib}
\bibliography{refs}

\begin{thebibliography}{10}

\bibitem{marzetta}
T.~L. Marzetta,
\newblock ``Noncooperative cellular wireless with unlimited numbers of base station antennas,''
\newblock {\em IEEE Transactions on Wireless Communications}, vol. 9, no. 11, pp. 3590--3600, 2010.

\bibitem{mmimo}
R.~C. de~Lamare,
\newblock ``Massive mimo systems: Signal processing challenges and future trends,''
\newblock {\em URSI Radio Science Bulletin}, vol. 2013, no. 347, pp. 8--20, 2013.

\bibitem{wence}
W.~Zhang, H.~Ren, C.~Pan, M.~Chen, R.~C. de~Lamare, B.~Du, and J.~Dai,
\newblock ``Large-scale antenna systems with ul/dl hardware mismatch: Achievable rates analysis and calibration,''
\newblock {\em IEEE Transactions on Communications}, vol. 63, no. 4, pp. 1216--1229, 2015.

\bibitem{ngo2015cell}
H.~Q. Ngo, A.~Ashikhmin, H.~Yang, E.~G. Larsson, and T.~L. Marzetta,
\newblock ``Cell-free massive mimo: Uniformly great service for everyone,''
\newblock IEEE 16th International Workshop on Signal Processing Advances in Wireless Communications (SPAWC), 2015.

\bibitem{interdonato2019ubiquitous}
G.~Interdonato, E.~Björnson, H.~Q. Ngo, P.~Frenger, and E.~G. Larsson,
\newblock ``Ubiquitous cell-free massive mimo communications,''
\newblock {\em EURASIP Journal on Wireless Communications and Networking}, vol. 2019, no. 1, pp. 1--13, 2019.

\bibitem{bjornson2020scalable}
E.~Bj{\"o}rnson and L.~Sanguinetti,
\newblock ``Scalable cell-free massive mimo systems,''
\newblock {\em IEEE Transactions on Communications}, vol. 68, no. 7, pp. 4247--4261, 2020.

\bibitem{llridd}
R.~B.~Di Renna and R.~C. de~Lamare,
\newblock ``Iterative detection and decoding with log-likelihood ratio based access point selection for cell-free mimo systems,''
\newblock {\em IEEE Transactions on Vehicular Technology}, vol. 73, no. 5, pp. 7418--7423, 2024.

\bibitem{ssettumba2024centralized}
T.~Ssettumba, Z.~Shao, L.~T.~N. Landau, M.~S.~P. Facina, P.~B. Da~Silva, and R.~C. De~Lamare,
\newblock ``Centralized and decentralized idd schemes for cell-free massive mimo systems: Ap selection and llr refinement,''
\newblock {\em IEEE Access}, 2024.

\bibitem{oclidd}
T.~Ssettumba, S.~Mashdour, L.~T.~N. Landau, P.~B. da~Silva, and R.~C. de~Lamare,
\newblock ``Iterative interference cancellation for clustered cell-free massive mimo networks,''
\newblock {\em IEEE Wireless Communications Letters}, pp. 1--1, 2024.

\bibitem{nayebi2016performance}
E.~Nayebi, A.~Ashikhmin, T.~L. Marzetta, and B.~D. Rao,
\newblock ``Performance of cell-free massive mimo systems with mmse and lsfd receivers,''
\newblock 2016 50th Asilomar Conference on Signals, Systems and Computers, 2016.

\bibitem{buzzi2017cell}
S.~Buzzi and C.~D’Andrea,
\newblock ``Cell-free massive mimo: User-centric approach,''
\newblock {\em IEEE Wireless Communications Letters}, vol. 6, no. 6, pp. 706--709, 2017.

\bibitem{tds}
Patrick Clarke and Rodrigo~C. de~Lamare,
\newblock ``Transmit diversity and relay selection algorithms for multirelay cooperative mimo systems,''
\newblock {\em IEEE Transactions on Vehicular Technology}, vol. 61, no. 3, pp. 1084--1098, 2012.

\bibitem{jpba}
Yuhan Jiang, Yulong Zou, Haiyan Guo, Theodoros~A. Tsiftsis, Manav~R. Bhatnagar, Rodrigo~C. de~Lamare, and Yu-Dong Yao,
\newblock ``Joint power and bandwidth allocation for energy-efficient heterogeneous cellular networks,''
\newblock {\em IEEE Transactions on Communications}, vol. 67, no. 9, pp. 6168--6178, 2019.

\bibitem{mashdour2024clustering}
S.~Mashdour, S.~Salehi, R.~C. de~Lamare, A.~Schmeink, and J.~P. S.~H. Lima,
\newblock ``Clustering and scheduling with fairness based on information rates for cell-free mimo networks,''
\newblock {\em IEEE Wireless Communications Letters}, 2024.

\bibitem{siprec}
Yunlong Cai, Rodrigo C.~de Lamare, and Rui Fa,
\newblock ``Switched interleaving techniques with limited feedback for interference mitigation in ds-cdma systems,''
\newblock {\em IEEE Transactions on Communications}, vol. 59, no. 7, pp. 1946--1956, 2011.

\bibitem{gbd}
Keke Zu, Rodrigo~C. de~Lamare, and Martin Haardt,
\newblock ``Generalized design of low-complexity block diagonalization type precoding algorithms for multiuser mimo systems,''
\newblock {\em IEEE Transactions on Communications}, vol. 61, no. 10, pp. 4232--4242, 2013.

\bibitem{wlbd}
Wence Zhang, Rodrigo~C. de~Lamare, Cunhua Pan, Ming Chen, Jianxin Dai, Bingyang Wu, and Xu~Bao,
\newblock ``Widely linear precoding for large-scale mimo with iqi: Algorithms and performance analysis,''
\newblock {\em IEEE Transactions on Wireless Communications}, vol. 16, no. 5, pp. 3298--3312, 2017.

\bibitem{mbthp}
Keke Zu, Rodrigo~C. de~Lamare, and Martin Haardt,
\newblock ``Multi-branch tomlinson-harashima precoding design for mu-mimo systems: Theory and algorithms,''
\newblock {\em IEEE Transactions on Communications}, vol. 62, no. 3, pp. 939--951, 2014.

\bibitem{robprec}
Y.~Cai, R.~C. de~Lamare, L.-L. Yang, and M.~Zhao,
\newblock ``Robust mmse precoding based on switched relaying and side information for multiuser mimo relay systems,''
\newblock {\em IEEE Transactions on Vehicular Technology}, vol. 64, no. 12, pp. 5677--5687, 2015.

\bibitem{bbprec}
Lukas T.~N. Landau and Rodrigo~C. de~Lamare,
\newblock ``Branch-and-bound precoding for multiuser mimo systems with 1-bit quantization,''
\newblock {\em IEEE Wireless Communications Letters}, vol. 6, no. 6, pp. 770--773, 2017.

\bibitem{rcfprec}
Victoria M.~T. Palhares, Andre~R. Flores, and Rodrigo~C. de~Lamare,
\newblock ``Robust mmse precoding and power allocation for cell-free massive mimo systems,''
\newblock {\em IEEE Transactions on Vehicular Technology}, vol. 70, no. 5, pp. 5115--5120, 2021.

\bibitem{lrcc}
Hang Ruan and Rodrigo~C. de~Lamare,
\newblock ``Distributed robust beamforming based on low-rank and cross-correlation techniques: Design and analysis,''
\newblock {\em IEEE Transactions on Signal Processing}, vol. 67, no. 24, pp. 6411--6423, 2019.

\bibitem{lrpa}
Tong Wang, Rodrigo~C. de~Lamare, and Anke Schmeink,
\newblock ``Joint linear receiver design and power allocation using alternating optimization algorithms for wireless sensor networks,''
\newblock {\em IEEE Transactions on Vehicular Technology}, vol. 61, no. 9, pp. 4129--4141, 2012.

\bibitem{rapa}
André~R. Flores and Rodrigo~C. de~Lamare,
\newblock ``Robust and adaptive power allocation techniques for rate splitting based mu-mimo systems,''
\newblock {\em IEEE Transactions on Communications}, vol. 70, no. 7, pp. 4656--4670, 2022.

\bibitem{rscf}
A.~R. Flores, R.~C. de~Lamare, and K.~V. Mishra,
\newblock ``Clustered cell-free multi-user multiple-antenna systems with rate-splitting: Precoder design and power allocation,''
\newblock {\em IEEE Transactions on Communications}, 2023.

\bibitem{demir2021foundations}
Ö.~T. Demir, E.~Björnson, and L.~Sanguinetti,
\newblock ``Foundations of user-centric cell-free massive mimo,''
\newblock {\em Foundations and Trends in Signal Processing}, vol. 14, no. 3-4, pp. 162--472, 2021.

\bibitem{d2021improving}
C.~D’Andrea and E.~G. Larsson,
\newblock ``Improving cell-free massive mimo by local per-bit soft detection,''
\newblock {\em IEEE Communications Letters}, vol. 25, no. 7, pp. 2400--2404, 2021.

\bibitem{mashdour2022MMSE-Based}
S.~Mashdour, R.~C. de~Lamare, A.~Schmeink, and J.~P. S.~H. Lima,
\newblock ``Mmse-based resource allocation for clustered cell-free massive mimo networks,''
\newblock presented at the IEEE 13th SCC, Braunschweig, Germany, 2023, pp. 149--154.

\bibitem{dimic2005downlink}
G.~Dimic and N.~D. Sidiropoulos,
\newblock ``On downlink beamforming with greedy user selection: performance analysis and a simple new algorithm,''
\newblock {\em IEEE Trans. Signal Process.}, vol. 53, no. 10, pp. 3857--3868, 2005.

\bibitem{van2016joint}
T.~Van~Chien, E.~Björnson, and E.~G. Larsson,
\newblock ``Joint power allocation and user association optimization for massive mimo systems,''
\newblock {\em IEEE Transactions on Wireless Communications}, vol. 15, no. 9, pp. 6384--6399, 2016.

\bibitem{bjornson2015optimal}
E.~Björnson, L.~Sanguinetti, J.~Hoydis, and M.~Debbah,
\newblock ``Optimal design of energy-efficient multi-user mimo systems: Is massive mimo the answer?,''
\newblock {\em IEEE Transactions on Wireless Communications}, vol. 14, no. 6, pp. 3059--3075, 2015.

\bibitem{bashar2018robust}
M.~Bashar, A.~Burr, K.~Haneda, and K.~Cumanan,
\newblock ``Robust user scheduling with cost 2100 channel model for massive mimo networks,''
\newblock {\em IET Microwaves, Antennas and Propagation}, vol. 12, no. 11, pp. 1786--1792, 2018.

\bibitem{chang2017energy}
Z.~Chang, Z.~Wang, X.~Guo, Z.~Han, and T.~Ristaniemi,
\newblock ``Energy-efficient resource allocation for wireless powered massive mimo system with imperfect csi,''
\newblock {\em IEEE Transactions on Green Communications and Networking}, vol. 1, no. 2, pp. 121--130, 2017.

\bibitem{rong2005robust}
Y.~Rong, S.~Shahbazpanahi, and A.~B. Gershman,
\newblock ``Robust linear receivers for space-time block coded multiaccess mimo systems with imperfect channel state information,''
\newblock {\em IEEE Transactions on Signal Processing}, vol. 53, no. 8, pp. 3081--3090, 2005.

\bibitem{vaezy2018energy}
H.~Vaezy, M.~J. Omidi, and H.~Yanikomeroglu,
\newblock ``Energy efficient precoder in multi-user mimo systems with imperfect channel state information,''
\newblock {\em IEEE Wireless Communications Letters}, vol. 8, no. 3, pp. 669--672, 2018.

\bibitem{rspa}
Saeed Mashdour, André~R. Flores, Shirin Salehi, Rodrigo~C. De~Lamare, Anke Schmeink, and Paulo~Branco Da~Silva,
\newblock ``Robust resource allocation in cell-free massive mimo systems,''
\newblock {\em IEEE Transactions on Communications}, pp. 1--1, 2025.

\bibitem{ngo2017cell}
H.~Q. Ngo, A.~Ashikhmin, H.~Yang, E.~G. Larsson, and T.~L. Marzetta,
\newblock ``Cell-free massive mimo versus small cells,''
\newblock {\em IEEE Transactions on Wireless Communications}, vol. 16, no. 3, pp. 1834--1850, 2017.

\bibitem{mashdour2022enhanced}
S.~Mashdour, R.~C. de~Lamare, and J.~P. S.~H. Lima,
\newblock ``Enhanced subset greedy multiuser scheduling in clustered cell-free massive mimo systems,''
\newblock {\em IEEE Communications Letters}, 2022.

\end{thebibliography}

\end{document}